\documentclass{ieeeaccess}
\usepackage{cite}
\usepackage{amsmath,amssymb,amsfonts}
\usepackage{algorithmic}
\usepackage{graphicx}
\usepackage{textcomp}
\usepackage{tabularx}
\usepackage{caption}
\usepackage{multirow}
\usepackage{booktabs}
\def\BibTeX{{\rm B\kern-.05em{\sc i\kern-.025em b}\kern-.08em
    T\kern-.1667em\lower.7ex\hbox{E}\kern-.125emX}}
\begin{document}
\history{Date of publication xxxx 00, 0000, date of current version xxxx 00, 0000.}
\doi{10.1109/ACCESS.2017.DOI}

\title{OcularAge: A Comparative Study of Iris and Periocular Images for Pediatric Age Estimation}
\author{\uppercase{Naveenkumar G Venkataswamy}\authorrefmark{1},
{\uppercase{ Poorna Ravi}}\authorrefmark{2},
\uppercase{Stephanie Schuckers\authorrefmark{3}, and Masudul H. Imtiaz,}.\authorrefmark{4}}
\address[1]{Department of Electrical and Computer Engineering, Clarkson University, New York, USA}
\address[2]{Department of Computer Science, Clarkson University, New York, USA}
\address[3]{Department of Electrical and Computer Engineering, Clarkson University, New York, USA}
\address[4]{Department of Electrical and Computer Engineering, Clarkson University, New York, USA}

\tfootnote{This material is based upon work supported by the Center for Identification Technology Research and the National Science Foundation under Grant No 1650503.}

\markboth
{Author \headeretal: Preparation of Papers for IEEE TRANSACTIONS and JOURNALS}
{Author \headeretal: Preparation of Papers for IEEE TRANSACTIONS and JOURNALS}

\corresp{Corresponding author: Masudul H. Imtiaz (mimtiaz@clarkson.edu).}

\begin{abstract}
Estimating a child's age from ocular biometric images is challenging due to subtle physiological changes and the limited availability of longitudinal datasets. Although most biometric age estimation studies have focused on facial features and adult subjects, pediatric-specific analysis, particularly of the iris and periocular regions, remains relatively unexplored. This study presents a comparative evaluation of iris and periocular images for estimating the ages of children aged between 4 and 16 years. We utilized a longitudinal dataset comprising more than 21,000 near-infrared (NIR) images, collected from 288 pediatric subjects over eight years using two different imaging sensors. A multi-task deep learning framework was employed to jointly perform age prediction and age-group classification, enabling a systematic exploration of how different convolutional neural network (CNN) architectures, particularly those adapted for non-square ocular inputs, capture the complex variability inherent in pediatric eye images. The results show that periocular models consistently outperform iris-based models, achieving a mean absolute error (MAE) of 1.33 years and an age-group classification accuracy of 83.82\%. These results mark the first demonstration that reliable age estimation is feasible from children’s ocular images, enabling privacy-preserving age checks in child-centric applications. This work establishes the first longitudinal benchmark for pediatric ocular age estimation, providing a foundation for designing robust, child-focused biometric systems. The developed models proved resilient across different imaging sensors, confirming their potential for real-world deployment. They also achieved inference speeds of less than 10 milliseconds per image on resource-constrained VR headsets, demonstrating their suitability for real-time applications.
\end{abstract}


\titlepgskip=-15pt

\maketitle

\section{Introduction}
With rising concerns around child safety in virtual environments, the need for accurate, privacy-preserving, and low-latency age estimation methods has become more urgent than ever, particularly as Virtual Reality (VR) headsets become increasingly accessible to younger audiences. Recently, estimating a person's age based on biometric traits has also gained significant traction in various application domains, including personalized healthcare, forensic analysis, and identity management systems~\cite{Ruth}. In particular, in VR environments, age estimation is critical in ensuring compliance with age-appropriate content delivery, protecting minors from explicit or unsafe experiences, and enabling adaptive user interfaces that respond to developmental needs~\cite{fiani2024exploring}. Here, biometric age estimation performed directly on-device could serve as a viable alternative to manual age input, which is often unreliable, easily bypassed, and prone to user disengagement.

While prior research has largely focused on facial features, ocular regions like the iris and periocular structures remain an underexplored yet promising avenue, especially for pediatric populations~\cite{Liao, Singh}. These regions are less susceptible to expression, occlusion, or pose variation, and remain accessible even when the lower face is masked or cropped. Recent studies have explored periocular regions for adult age estimation. For example, Bisogni et al.~\cite{bisogni2022periocular} achieved 84\% accuracy using periocular RGB videos for age-group classification. However, pediatric ocular age estimation, particularly using NIR imagery, remains largely unaddressed. More importantly, periocular and iris features encode age-related cues such as eyelid geometry, scleral reflectance, and iris texture\cite{fairhurst2011analysis}, which evolve over time. Despite this potential, their use in pediatric populations remains underexplored, and no prior work has systematically compared the predictive value of iris versus periocular regions in children. Moreover, no prior study has conducted this comparison under realistic deployment conditions, such as cross-sensor settings and latency-constrained hardware. In this study, we address this gap by leveraging a longitudinal dataset of over 21,000 NIR ocular images collected from 288 subjects aged 4 to 16, and evaluating age estimation performance across both modalities using deep learning models optimized for real-world deployment.

Ocular biometrics offer several practical advantages: they are non-invasive, generally stable over time, and valuable in cases where face images may be unavailable or unsuitable~\cite{Mehrotra}. Despite this, most prior work on age estimation has concentrated on adult datasets~\cite{Erbilek, odion2022age}, leaving open questions about how age-related changes manifest in children’s ocular features. As children grow, ocular anatomy undergoes significant changes, shaped by both biological maturation and environmental exposure. Prior studies have shown that characteristics like axial length, corneal curvature, and iris texture exhibit measurable growth during childhood and adolescence~\cite{Hashemi, Zengin, Read}. Such developmental variability poses challenges for models trained primarily on adult datasets, which may fail to generalize across pediatric age groups. 

One fundamental question that remains is which ocular region offers the most informative features for pediatric age prediction. Some studies rely solely on iris-based features~\cite{Erbilek, sgroi2013prediction}, while others have explored facial analysis~\cite{Alonso}, but a systematic comparison of ocular sub-regions in children remains lacking. Including broader contextual cues, such as eyelid geometry, scleral patterns, and periocular skin texture, may enhance performance; however, this has not been thoroughly explored.

Another essential consideration is sensor variation. In practical settings, images are often captured with different devices, each introducing variations in resolution, lighting, and image quality. This sensor-related variability, known as sensor bias, can affect model generalization~\cite{Gangwar}. Although sensor interoperability has been studied in other biometric modalities, its impact on pediatric ocular age estimation remains underexplored.

Finally, for real-time applications such as virtual reality, models must be lightweight and responsive. VR environments require low-latency inference to maintain user comfort~\cite{Sagnier}, placing strict constraints on model complexity and processing time. In this work, we evaluate multi-task deep learning models (for age group classification and exact age regression) trained on both iris and periocular images, compare their performance across age groups and sensor conditions, and assess their deployment viability on resource-constrained VR hardware.

Our main contributions are as follows:
\begin{itemize}
    \item A comparative analysis of iris and periocular (eye-based) images using a multi-task learning framework for both age-group classification and exact age regression, demonstrating the advantages of combining coarse and fine-grained predictions.
    \item A multi-task learning framework trained on subject-exclusive splits, incorporating soft labels and confidence-based age analysis across developmental stages to interpret model uncertainty and age-specific variations.
    \item A demonstration of real-time feasibility on the Oculus Quest 2 headset, deploying MobileNetV3 with inference latency suitable for immersive, privacy preserving on-device applications.
\end{itemize}

The remainder of the paper is organized as follows. Section~\ref{sec:related_work} reviews relevant work in biometric age estimation and ocular analysis. Section~\ref{sec:methods} describes our dataset, pre-processing steps, and model training procedures. Section~\ref{sec:results} presents experimental results, sensor generalization, and VR deployment benchmarks. Section~\ref{sec:discussion} outlines limitations and broader implications. Section~\ref{sec:conclusion} concludes the paper.

\section{Background and Related Work}
\label{sec:related_work}

The rapid expansion of VR platforms among minors has prompted urgent questions around digital safety and age-appropriate access. In this context, biometric-based, on-device age estimation presents a timely and privacy-conscious alternative to conventional age verification methods. Age verification in VR platforms is currently dominated by third-party identity verification services that combine government issued IDs with biometric checks~\cite{vrchat}. These systems, such as those offered by Persona~\cite{persona}, prompt users to submit identification documents via smartphone or webcam, from which relevant metadata, such as date of birth, is extracted and processed. While effective for formal verification, this approach introduces onboarding friction, depends on external infrastructure, and is not optimized for real-time, on-device use in child-centric applications.

Ocular biometrics, particularly in pediatric populations, present a promising alternative. Prior research has shown that ocular structures undergo measurable physiological changes throughout childhood. Hashemi et al.~\cite{Hashemi} reported a strong correlation between age and axial length, noting an annual increase of 0.16 to 0.39 mm among children aged 6 to 18. Similarly, Zengin et al.~\cite{Zengin} reported age-dependent changes in choroidal thickness across subjects aged 4 to 23. These findings indicate that ocular anatomy contains age-related indicators, rendering it a feasible method for estimating biological age.

Yet, most iris-focused research has concentrated on identity recognition rather than age estimation. Recent longitudinal studies have explored how iris stability holds up as children grow, offering new insights into its persistence during early development. Das et al.~\cite{das2021iris} studied data from 209 children aged 4 to 11 over three years, revealing that although match scores exhibited a slight decline over time, the impact of aging was statistically significant but practically minor. A follow-up study by Das et al.~\cite{das2023longitudinal}, spanning up to 6.5 years, confirmed that iris recognition systems maintained high accuracy for young individuals, with no operational degradation over time. These findings suggest that iris recognition remains reliable in pediatric populations, with match score variability mainly attributed to factors such as pupil dilation rather than structural changes.

Strong recognition performance doesn’t necessarily mean that age related biometric signals are absent. Fairhurst and Erbilek~\cite{fairhurst2011analysis} demonstrated that natural physiological changes in the iris due to aging can lead to measurable structural variation, even if these do not compromise identity matching. This suggests that while recognition systems may remain robust, the same subtle structural cues, such as changes in iris texture, pupil boundary, or stroma patterns, may still carry predictive information about age.

Only a few studies have investigated age prediction directly from iris imagery. Erbilek et al.\cite{Erbilek} used geometric features and support vector machines to classify adult age groups, achieving 75\% accuracy on a limited dataset. Another study by Sgroi et al.\cite{sgroi2013prediction} introduced an iris texture-based classification approach that distinguishes between older and younger subjects by exploiting subtle, age-related textural cues present in the iris. More recently, deep learning approaches have shown promise in modeling subtle ocular changes. Rajput et. al.~\cite{rajput2019deep} studies an iris-based gender and age classification, indicating that deep CNNs can capture information acroos the iris that is indicative of both gender and aging cues.Despite these advances, most studies remain adult-focused, with children either underrepresented or analyzed alongside adults, limiting insights specific to early developmental stages. This is particularly critical, as ocular development accelerates during puberty\cite{Hansen}, introducing more complex and nonlinear changes.

Sensor variability further complicates the modeling of ocular traits. Differences in imaging devices—such as resolution, illumination, and optical distortion can affect both segmentation and downstream performance. Gangwar and Joshi~\cite{Gangwar} reported a drop in iris recognition accuracy under cross-sensor evaluation, emphasizing the sensitivity of deep models to acquisition conditions. This issue is particularly pronounced in pediatric contexts, where physiological responses to environmental stimuli differ significantly from those of adults. For instance, Hartstein et al.~\cite{Hartstein} observed that children exhibit substantially faster and stronger pupillary constriction in response to blue light compared to adolescents, indicating sensor response differences that could affect feature consistency.

Real-time deployment introduces additional constraints. VR platforms, in particular, demand `sub-11 ms inference latency' to preserve user comfort~\cite{meta}. High-capacity models often exceed this threshold, limiting their practical use on standalone systems. Despite these challenges, no prior study has systematically examined pediatric age estimation using both iris and periocular modalities across modern deep learning backbones, nor assessed their suitability for real-time deployment in constrained VR environments.

Our work addresses these gaps by presenting the first systematic comparison of pediatric age estimation using two ocular input modalities: normalized iris images and periocular eye images. We benchmark six deep learning architectures—EfficientNet-B3~\cite{alhichri2021classification}, MobileNetV3~\cite{koonce2021mobilenetv3}, ResNet-50~\cite{koonce2021resnet}, DenseNet-121~\cite{Huang}, ConvNeXt-Tiny~\cite{liu2022convnet}, and the hybrid MobileViT~\cite{mehta2021mobilevit}—across both input types. Further, we evaluate their generalization performance under cross-sensor conditions and assess their feasibility for deployment on VR platforms. These experiments aim to establish practical and scalable baselines for real-time pediatric age estimation using ocular biometrics.

\section{Methodology}
\label{sec:methods}
\subsection{Dataset Preparation}

\subsubsection{Data Collection}

The dataset used in this study was collected as part of a longitudinal biometric research effort~\cite{das2021iris, das2023longitudinal} conducted at Clarkson University, with a specific focus on studying biometric recognition in children. Although the broader study involved multiple biometric modalities, this work focuses exclusively on the iris modality.

Participants were recruited from public schools in the Potsdam, NY area, including elementary, middle, and high schools. Participation was voluntary, with informed consent obtained from parents and age-appropriate assent from the children. All procedures were approved by Clarkson University’s Institutional Review Board (IRB).

Initial enrollment included children aged 4 to 11. Over time, the study expanded to include younger children in Pre-K, typically aged 4 to 5. Data collection occurred twice annually, in the spring and fall, allowing for roughly six-month intervals between sessions. This biannual schedule was designed to support longitudinal analysis of developmental changes in ocular biometrics. Four collection sessions were canceled due to the COVID-19 pandemic, resulting in occasional gaps in the timeline and reducing temporal continuity for some participants.

\begin{figure*}
    \centering
    \includegraphics[width=\linewidth]{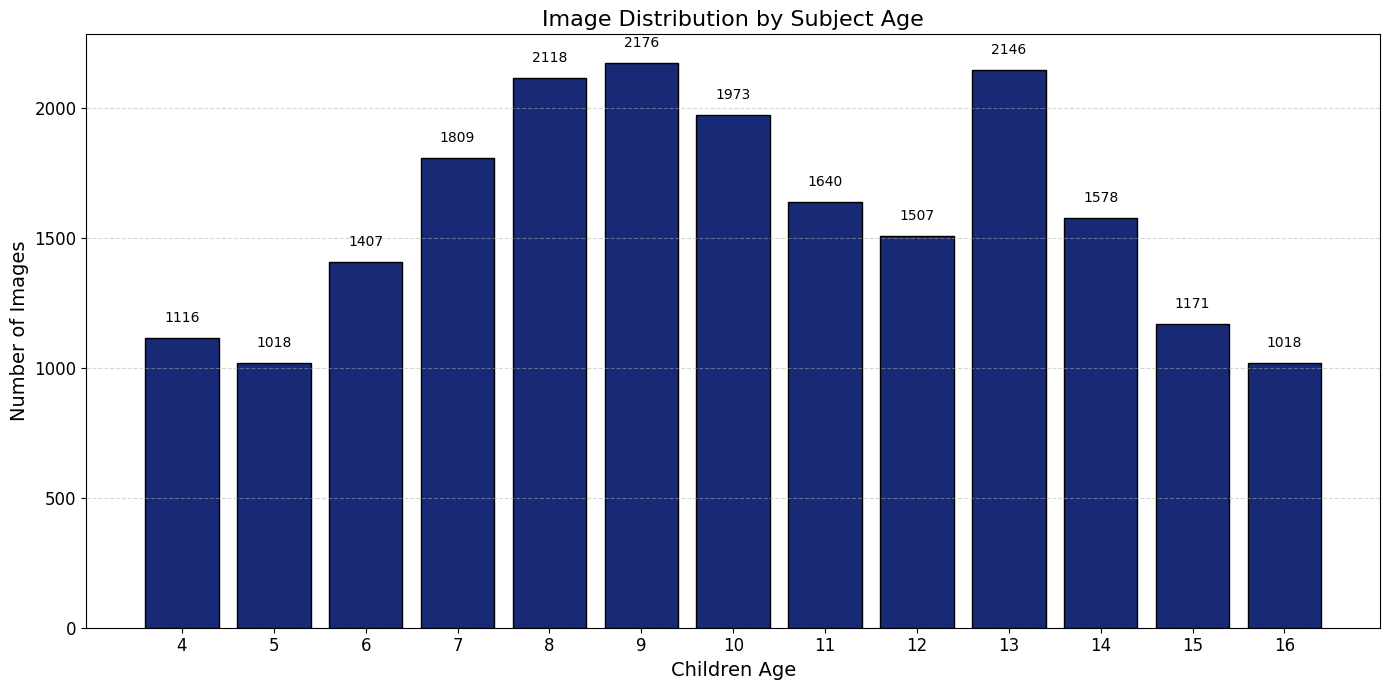}
    \caption{Distribution of images by subject age.}
    \label{fig:age count}
\end{figure*}

As is common in long-term pediatric studies, variation in individual participation occurred due to absences, relocation, or withdrawal from the study. Consequently, not all participants contributed data at every time point. The dataset comprises over 21,000 ocular images from 288 unique subjects, aged 4 to 16, collected over eight years. Figure~\ref{fig:age count} shows the distribution of images across subject age, with both left and right eye samples considered.

\begin{figure}
    \centering
    \includegraphics[width=\linewidth]{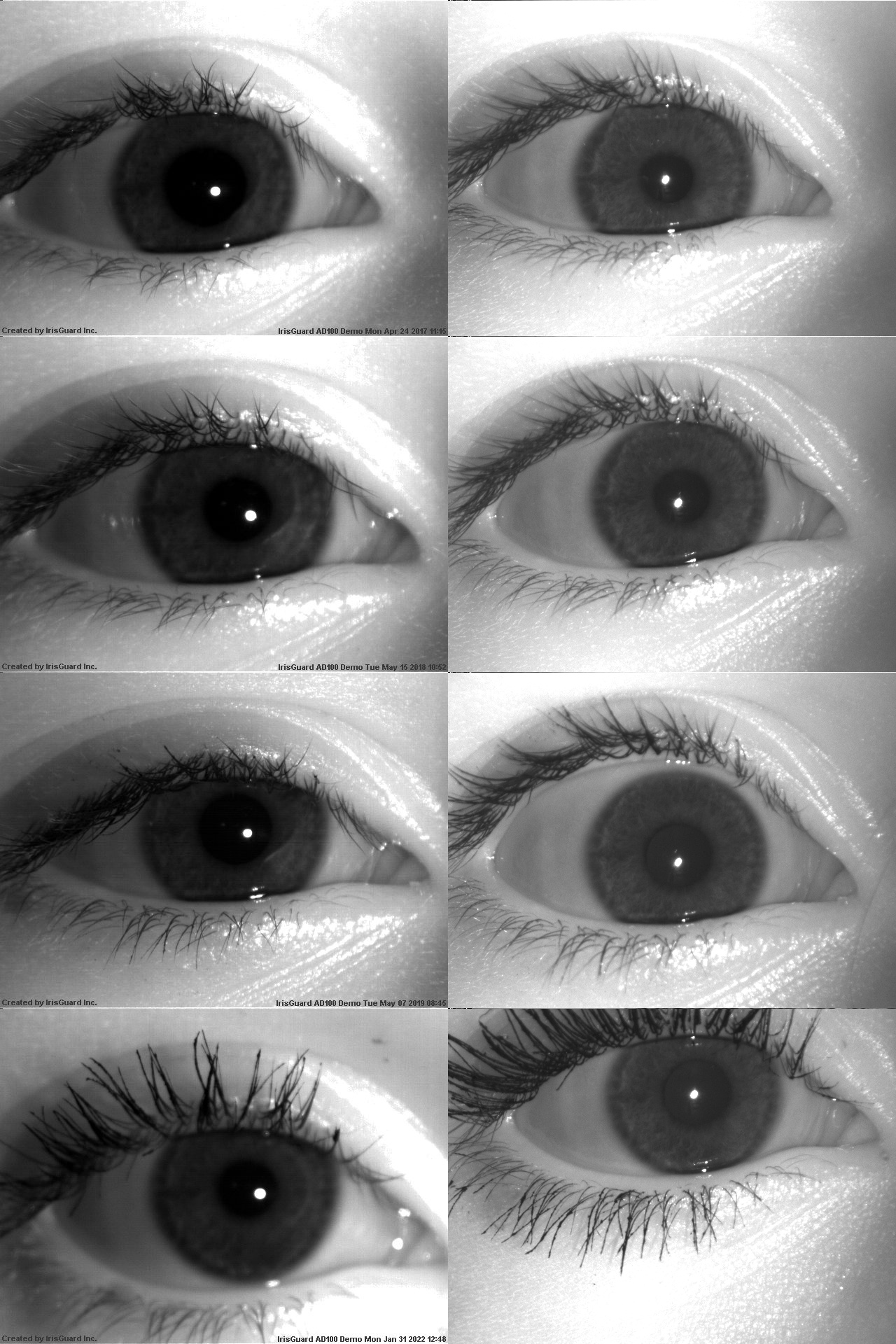}
    \caption{Example NIR eye images captured using two different iris imaging devices. The left column shows images collected with the IG-AD100, while the right column shows images from the iCAM T10. Rows correspond to ages 9, 10, 11, and 12 (top to bottom), illustrating consistent cross-sensor capture and longitudinal variation.}
    \label{fig:both_scanner_eye}
\end{figure}

Two NIR iris imaging devices were used throughout the study. The IrisGuard IG-AD100 Dual Iris Camera (IG-AD100) ~\cite{irisguard} was used in earlier phases and accounted for the majority of sessions, yielding 16,285 images. The Iris ID iCAM T10 (iCAM T10)~\cite{irisID} was introduced in later sessions, contributing an additional 5,014 images. Representative samples from both devices are shown in Figure~\ref{fig:both_scanner_eye}.

In terms of sensor distribution, 16,885 eye images were captured using the IG-AD100, while 5,337 eye images were collected with the iCAM T10. Although the IG-AD100 was used for most sessions, each age group includes data from both sensors. A summary of the dataset composition is provided in Table~\ref{tab:dataset}.

Data collection was carried out under semi-controlled lighting conditions to maintain consistent image quality. Blinds were drawn to minimize interference from ambient NIR light, and participants were given time to adapt to the lighting environment to reduce pupil dilation artifacts.

Multiple studies have been performed on this iris dataset, including~\cite{das2021iris, das2023longitudinal}, which investigated the longitudinal stability of iris recognition in children. The first study analyzed data from 209 subjects aged 4 to 11 over 3 years, reporting a statistically significant but practically negligible aging effect~\cite{das2021iris}. The follow-up study extended the analysis to 230 subjects across 6.5 years, confirming the long-term viability of iris recognition in pediatric populations and concluding no measurable degradation in matching performance due to aging~\cite{das2023longitudinal}.

\begin{table}[!t]
\centering
\caption{Summary of the final dataset used for training and evaluation. The table displays the number of preprocessed eye and iris images, categorized by imaging device and age group. While the IG-AD100 contributed the majority of images, both age groups included samples from both sensors.}
\label{tab:dataset}
\begin{tabular}{|l|c|c|}
\hline
\textbf{Category} & \textbf{Eye Images} & \textbf{Iris Images} \\
\hline
Total images & 21,922 & 21,619 \\
\hline
Total IG-AD100 & 16,585 & 16,422 \\
\hline
Total iCAM T10 & 5,337 & 5,197 \\
\hline
Age group 1 (4--9) & 11,096 & 10,936 \\
\hline
Age group 2 (10--16) & 10,826 & 10,683 \\
\hline
\end{tabular}
\end{table}

\subsubsection{Data Pre-processing}
To prepare the data for model training, we first calculated the age of each participant by subtracting their birth year from the year the image was captured. The dataset spans approximately 17 biannual collection sessions conducted over eight years, with four sessions missed due to the COVID-19 pandemic. This resulted in a subject age range from 4 to 16 years.

To better capture developmental variations, we grouped the data into two age categories:
\begin{itemize}
    \item \textbf{4--9 years:} Early childhood, a phase characterized by rapid ocular development. This group includes 11,096 images from 221 participants.
    \item \textbf{10--16 years:} Middle childhood through early adolescence, marked by moderate and more variable ocular changes. This group contains 10,826 images from 222 participants.
\end{itemize}

Both left and right eye images were included to increase diversity and improve model generalization. Although the eyes of a subject are naturally correlated, we maintained a subject-exclusive split across all partitions to avoid data leakage and overfitting. Specifically, \textbf{all images from a given subject were assigned to only one of the training (80\%), validation (10\%), or test (10\%) sets, ensuring no overlap across splits}.

Original eye images, captured at a resolution of $640\times480$ pixels, were resized to $320\times240$ pixels. No explicit segmentation or masking was applied to periocular images. To enhance robustness under real-world imaging conditions, we applied a range of data augmentation techniques, including random rotations, affine transformations, horizontal flipping, and contrast variation. These augmentations simulate variability in head pose, lighting, and capture conditions.

For iris-specific pre-processing, we used the OSIRIS toolkit~\cite{othman2016osiris} to segment and normalize the iris region from each eye image. The segmentation process includes detecting iris and pupil boundaries, identifying eyelid occlusions using active contour models, and normalizing the extracted iris region to a $256\times32$ rectangular strip using Daugman’s rubber sheet model ~\cite{daugman2009iris}. OSIRIS also produces a binary mask for each image, where pixels corresponding to valid iris regions are marked as 1, and occluded or padded areas are set to 0.

\begin{figure*}
    \centering
    \includegraphics[width=\linewidth]{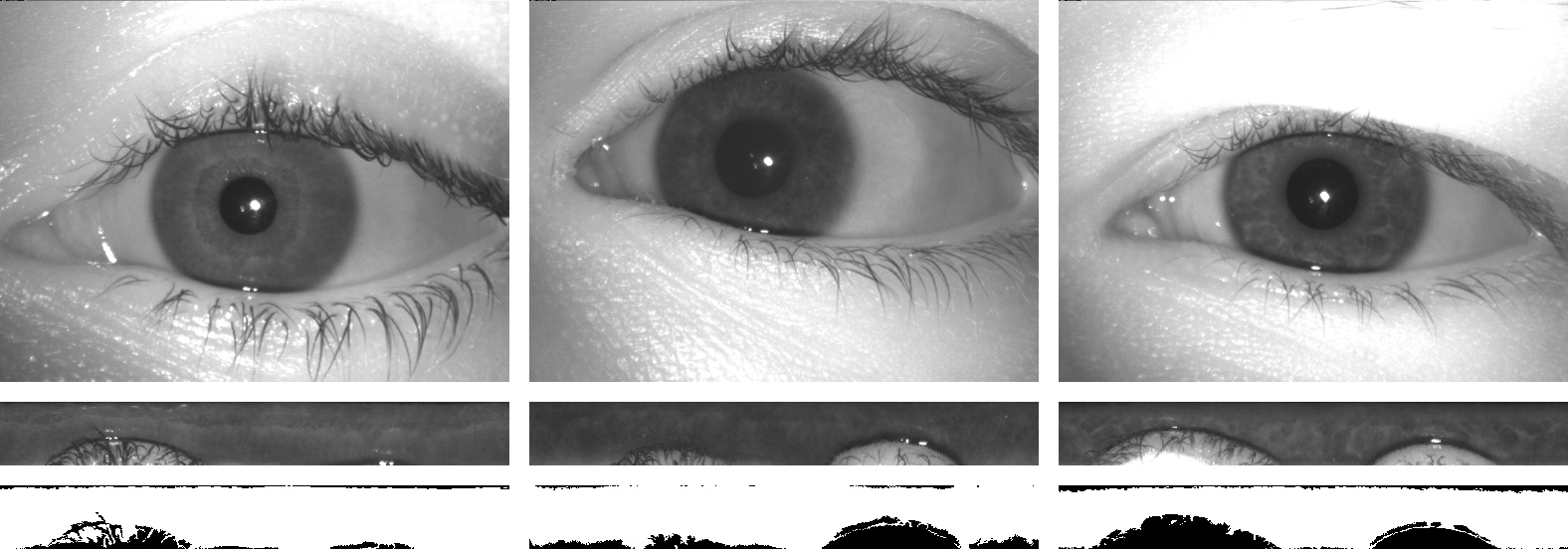}
    \caption{Example of the iris pre-processing pipeline. From top to bottom: the original eye image, the corresponding normalized iris image generated using OSIRIS, and the binary mask indicating the valid iris region. The mask is used as a second channel during model training to guide the network’s attention to meaningful iris features.}
    \label{fig:eye_iris_mask}
\end{figure*}

Figure~\ref{fig:eye_iris_mask} illustrates a representative example of this pre-processing workflow. The top image shows the original eye input, the middle shows the normalized iris region, and the bottom displays the binary mask highlighting valid iris pixels. During training, this binary mask was concatenated with the 1-channel grayscale image as a second input channel, allowing the model to focus on valid iris regions and ignore irrelevant or occluded areas, improving both interpretability and robustness.

Approximately 300 images were excluded due to segmentation failures. After filtering, we retained 16,422 iris images from the IG-AD100 device and 5,197 iris images from the iCAM T10. The summary of all preprocessed images is shown in Table~\ref{tab:dataset}. Additional augmentations applied to iris data included random Gaussian blur and intensity normalization to improve generalization and reduce overfitting. The lower count of iris images compared to eye images is attributed to segmentation dropouts during pre-processing.

\subsection{Model Selection}

To benchmark age estimation from ocular images, we selected a focused set of deep learning models that support rectangular input resolutions and are suitable for deployment on resource-constrained hardware. This set includes lightweight and mid-sized CNNs, as well as hybrid architectures such as MobileViT that incorporate spatial attention mechanisms. The selection was motivated by the need to preserve the natural aspect ratio of eye and iris images while balancing the extraction of fine-grained texture features with computational efficiency.

The following six architectures were chosen to cover a range of model complexities and design philosophies:
\begin{itemize}
    \item \textbf{EfficientNet-B3}~\cite{alhichri2021classification}, which applies compound scaling to balance network depth, width, and input resolution, improving accuracy with a relatively modest computational footprint.
    \item \textbf{MobileNetV3}~\cite{koonce2021mobilenetv3}, designed for efficiency on edge devices, using depthwise separable convolutions and lightweight nonlinearities.
    \item \textbf{DenseNet-121}~\cite{Huang}, which employs densely connected layers to encourage feature reuse and alleviate vanishing gradients in deeper networks.
    \item \textbf{ResNet-50}~\cite{koonce2021resnet}, a widely used baseline that incorporates residual connections for stable training and improved convergence in deeper architectures.
    \item \textbf{ConvNeXt-Tiny}~\cite{liu2022convnet}, which blends design elements from Transformers into a CNN style hierarchy, offering strong feature representation with modern architectural advances.
    \item \textbf{MobileViT-S}~\cite{mehta2021mobilevit}, a hybrid model combining convolutional layers with lightweight attention blocks, enabling both local detail extraction and global spatial context modeling.
\end{itemize}

These models span a broad spectrum of trade-offs. Architectures such as MobileNetV3 and MobileViT-S are optimized for low latency and minimal memory usage, making them ideal for real-time deployment on devices like the Oculus Quest 2. In contrast, larger models such as ResNet-50 and EfficientNet-B3 offer greater representational capacity at the cost of increased computational demand.

To assess both recognition accuracy and deployment feasibility, each model was evaluated on two input modalities: iris and eye images. This dual-modality evaluation allowed us to examine how architectural characteristics interact with the input context. A summary of the models used, computational cost, and average inference time (measured on an NVIDIA TITAN RTX GPU) is provided in Table~\ref{tab:modelstats}.

\begin{table}[!t]
\centering
\caption{Comparison of selected models by number of parameters, Floating Point Operations (FLOPs), and average inference time measured on an NVIDIA TITAN RTX GPU. FLOPs are computed per forward pass on a $320\times240$ input image.}
\label{tab:modelstats}
\begin{tabularx}{\linewidth}{|l|X|X|X|}
\hline
\textbf{Model} & \textbf{Parameters (M)} & \textbf{FLOPs (B)} & \textbf{Inference Time (ms)} \\
\hline
ResNet-50       & 25.6   & 4.1   & 0.15 \\
\hline
EfficientNet-B3 & 12.0   & 1.8   & 0.10 \\
\hline
DenseNet-121    & 8.0    & 2.9   & 0.12 \\
\hline
MobileNetV3     & 5.4    & 0.2   & 0.07 \\
\hline
ConvNeXt-Tiny   & 28.6   & 4.5   & 0.18 \\
\hline
MobileViT-S     & 5.6    & 1.1   & 0.08 \\
\hline
\end{tabularx}
\end{table}

\subsection{Training Configuration}

\begin{figure*}[!t]
    \centering
    \includegraphics[width=\linewidth]{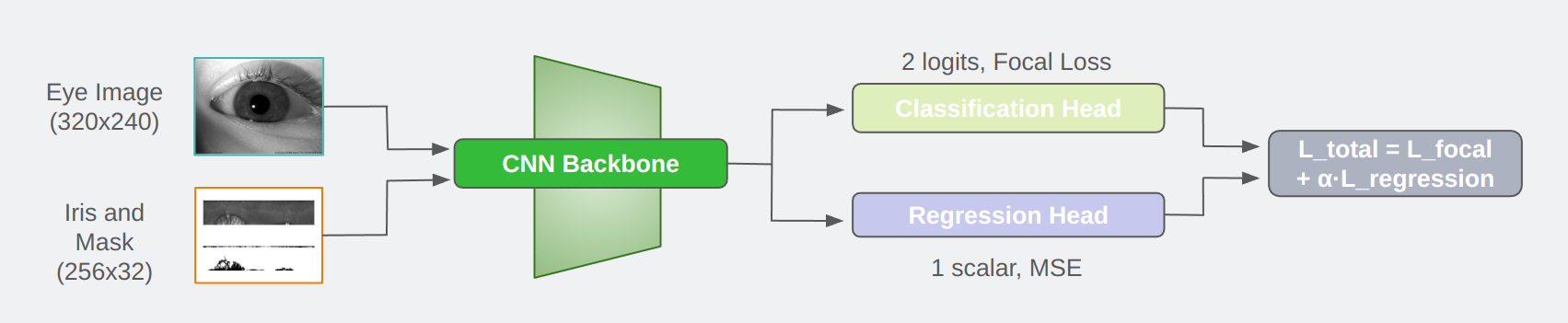}
    \caption{Overview of the model architecture. Eye-based models take a 1-channel grayscale input, while iris-based models use a 2-channel input comprising the normalized iris image and its binary mask. All models share a modified input stem adapted from pretrained ImageNet backbones, followed by a shared feature extractor (backbone) and two output heads: one for age group classification and one for exact age regression.}
    \label{fig:model_architecture}
\end{figure*}

\subsubsection{Training Objectives}
We formulate pediatric age estimation as a multi-task learning problem with two outputs: a binary classification head for age group prediction (4--9 vs. 10--16 years), and a regression head for estimating exact age. This dual task formulation captures both coarse categorical distinctions and fine-grained age progression. Each model jointly optimizes both objectives during training.

\subsubsection{Model Initialization and Architecture Modifications}
All models were initialized using ImageNet pretrained weights. The final classification layers were replaced with a dual output structure: (1) a fully connected classification head with two logits, and (2) a regression head that outputs a single scalar value. As shown in Figure~\ref{fig:model_architecture}, eye-based models receive grayscale images as 1-channel input, while iris-based models use two channels, one for the normalized iris image and one for the corresponding binary mask. To accommodate these variations, the input stem of each network was modified by averaging pretrained RGB weights for grayscale inputs and adjusting the first convolutional layer to handle two channels for iris models.

\subsubsection{Loss Functions and Multi-Task Optimization}
For classification, we used sigmoid focal loss to address the class imbalance and mitigate overconfidence near age group boundaries. Mean squared error (MSE) loss was used for the regression task. The overall training objective was defined as:

\[
\mathcal{L}_{\text{total}} = \mathcal{L}_{\text{focal}} + \alpha \cdot \mathcal{L}_{\text{regression}}
\]

where $\alpha$ adjusts the relative contribution of the regression loss. We initialized $\alpha$ to 0.25 and dynamically adjusted it during training to maintain proportional contributions from both tasks.

\subsubsection{Training Procedure and Hyper-parameters}

All models were trained for up to 50 epochs using the Adam optimizer, with an initial learning rate of 0.001, weight decay of 0.01, and a batch size of 64. Cosine annealing with warm restarts was used as the learning rate schedule. Early stopping based on validation loss was applied, with a patience threshold of 5 epochs. To address the class imbalance, we incorporated inverse-frequency class weighting during the computation of classification loss. A complete summary of the core training hyper-parameters used across all models is provided in Table~\ref{tab:training_hparams}.

\subsubsection{Augmentation and Regularization}
To improve generalization, we applied both spatial and photometric augmentations during training, including random horizontal flipping, affine transformations (rotation, translation, scaling), Gaussian blur, sharpness adjustment, and auto contrast. Label smoothing was used on the classification head to regularize overconfident predictions, particularly near the class transition boundary.

\subsubsection{Implementation Details and Reproducibility}

All experiments were implemented using PyTorch 2.1.0 and CUDA 12.1. The training was conducted on an NVIDIA TITAN RTX GPU with 24~GB of VRAM. To ensure reproducibility, we set the random seeds for PyTorch and NumPy to 42. Models followed their original architectural normalization layers using batch normalization for CNNs and layer normalization for hybrid models such as MobileViT-S.

All input images were normalized using dataset-specific mean and standard deviation values. For eye-based models, we used $\texttt{mean} = [0.5187]$ and $\texttt{std} = [0.2505]$. For iris-based models, we used $\texttt{mean} = [0.2103]$ and $\texttt{std} = [0.0879]$. These values were computed from the respective training sets and used consistently across all experiments.

\begin{table*}[h]
\centering
\caption{Core Training hyper-parameters Used Across All Models}
\label{tab:training_hparams}
\begin{tabular}{|l|c|}
\hline
\textbf{Hyperparameter} & \textbf{Value} \\
\hline
Normalization Mean (Eye) & 0.5187 \\
\hline
Normalization Standard Deviation (Eye) & 0.2505 \\
\hline
Normalization Mean (Iris) & 0.2103 \\
\hline
Normalization Standard Deviation (Iris) & 0.0879 \\
\hline
Batch Size & 64 \\
\hline
Epochs & 50 \\
\hline
Optimizer & Adam \\
\hline
Initial Learning Rate & 0.001 \\
\hline
Weight Decay & 0.01 \\
\hline
Learning Rate Scheduler & Cosine Annealing with Warm Restarts \\
\hline
Early Stopping Patience & 5 epochs \\
\hline
Loss Functions & Sigmoid Focal (classification), MSE (regression) \\
\hline
Loss Weight ($\alpha$) & 0.25 (dynamically adjusted) \\
\hline
Input Size (Eye) & $320\times240$ \\
\hline
Input Size (Iris) & $256\times32$ (with binary mask as second channel) \\
\hline
Label Smoothing & Applied (classification) \\
\hline
Class Weighting & Inverse-frequency weighting \\
\hline
Augmentations & Flip, affine, blur, sharpness, auto contrast \\
\hline
Random Seed & 42 \\
\hline
Framework & PyTorch 2.1.0, CUDA 12.1 \\
\hline
Hardware & NVIDIA TITAN RTX, 24GB VRAM \\
\hline
\end{tabular}
\end{table*}

\subsection{Model Evaluation and Deployment Benchmarking}

To evaluate model performance across architectures and input types, we conducted a comprehensive benchmarking study. Each model was trained independently on both iris and eye images using subject-exclusive splits. The evaluation was performed on a held-out test set comprising participants not seen during training or validation.

The multi-task framework outputs two predictions per image: (1) a binary classification of age group (4--9 vs. 10--16 years), and (2) a regression estimate of exact age. Classification performance was assessed using overall accuracy, per-class precision, recall, and F1-score. For regression, we computed MAE, RMSE, and accuracy within $\pm$1 and $\pm$2 years of the ground truth age. To analyze robustness across developmental stages, we further evaluated performance across four age bins: 4--6, 7--9, 10--12, and 13--16 years.

To ensure fair comparison, all models were trained using identical data splits, augmentation pipelines, and training configurations. The evaluation was conducted separately for eye-based and iris-based models to assess the predictive utility of each modality.

In addition to accuracy metrics, we evaluated the deployment feasibility of each model across three hardware platforms: an NVIDIA TITAN RTX GPU, a Jetson AGX Orin embedded system, and the Oculus Quest 2 headset. This multi-platform assessment enables direct comparison between high-performance and resource-constrained settings.

\textbf{GPU-based Evaluation:} All models were initially trained and benchmarked on an NVIDIA TITAN RTX GPU using PyTorch. This served as the baseline environment for accuracy and inference speed. Only forward pass latency was measured, excluding pre-processing overhead. Metrics for both classification and regression tasks were computed in aggregate and per class.

\begin{table*}[h]
\centering
\caption{Overall classification and regression performance using eye images.}
\label{tab:eye_overall_results}
\begin{tabular}{|l|c|c|c|c|c|}
\hline
\textbf{Model} & \textbf{Accuracy (\%)} & \textbf{MAE} & \textbf{RMSE} & \textbf{F1-Score (Avg)} & \textbf{$\pm$1yr Acc (\%)} \\
\hline
MobileNetV3-Large  & \textbf{83.82} & \textbf{1.33} & \textbf{1.82} & \textbf{0.83} & \textbf{84.56} \\
\hline
EfficientNet-B3    & 80.58 & 1.33 & 1.81 & 0.80 & 74.71 \\
\hline
DenseNet-121       & 80.22 & 1.54 & 2.03 & 0.77 & 72.19 \\
\hline
ResNet-50          & 77.12 & 1.62 & 2.09 & 0.76 & 70.98 \\
\hline
MobileViT-S        & 77.12 & 1.55 & 2.00 & 0.74 & 75.65 \\
\hline
ConvNeXt-Tiny      & 45.23 & 2.53 & 3.16 & 0.00 & 0.00 \\
\hline
\end{tabular}
\end{table*}

\textbf{Edge Deployment on Jetson AGX Orin:} For embedded deployment, models were ported to the Jetson AGX Orin platform (32~GB RAM, 2048-core GPU). After full-precision (FP32) training, we fine-tuned the models for several epochs using PyTorch’s Automatic Mixed Precision (AMP) to approximate lower precision behavior. The trained models were exported to ONNX format and optimized using TensorRT with post-training quantization, including dynamic range calibration and bias correction. This pipeline preserved near FP32 accuracy while significantly reducing inference time. Latency was measured over batches of 1,000 images under sustained execution.

\textbf{VR Deployment on Oculus Quest 2:} The MobileNetV3 model was selected for deployment on the Oculus Quest 2 (Qualcomm Snapdragon XR2). Following FP32 training and AMP-based fine-tuning, the model was exported to ONNX, converted to TensorFlow format, and quantized using TensorFlow Lite (TFLite) with FP16 precision. The quantized model was deployed in Unity (2022.3 LTS) using the Barracuda inference engine. Inference latency was measured over 1,000 images using the Oculus Performance Toolkit. MobileNetV3 was selected due to its balance of accuracy and efficiency, making it well-suited for real-time inference in immersive virtual reality (VR) environments.

\subsection{Sensor Bias Analysis}

Sensor bias can significantly affect the generalization ability of deep learning models, particularly when training and testing occur across different imaging devices. These differences may arise from variations in resolution, illumination conditions, sensor noise, optical characteristics, and other device-specific parameters. In this study, two NIR imaging devices were used: the IG-AD100~\cite{irisguard}, which accounted for the majority of the dataset, and the iCAM T10~\cite{irisID}, introduced in the later stages of data collection.

To investigate the impact of sensor variability, we conducted cross-sensor experiments in which models were trained exclusively on data from the IG-AD100 and evaluated on samples acquired using both IG-AD100 and the iCAM T10. This setup allowed us to assess model generalization to previously unseen sensor characteristics under real-world deployment conditions.
\begin{table*}[h]
\centering
\caption{Per-class classification metrics (Class 0: 4--9, Class 1: 10--16).}
\label{tab:eye_classwise_metrics}
\begin{tabular}{|l|c|c|c|}
\hline
\textbf{Model} & \textbf{Precision (C0 / C1)} & \textbf{Recall (C0 / C1)} & \textbf{F1 (C0 / C1)} \\
\hline
MobileNetV3-Large  & 0.85 / 0.82 & 0.83 / 0.82 & 0.82 / 0.81 \\
\hline
EfficientNet-B3    & 0.84 / 0.77 & 0.80 / 0.81 & 0.82 / 0.79 \\
\hline
DenseNet-121       & 0.80 / 0.78 & 0.83 / 0.75 & 0.82 / 0.77 \\
\hline
ResNet-50          & 0.82 / 0.72 & 0.75 / 0.80 & 0.78 / 0.76 \\
\hline
MobileViT-S        & 0.78 / 0.76 & 0.81 / 0.73 & 0.80 / 0.74 \\
\hline
\end{tabular}
\end{table*}

Subject-exclusive splits were used to prevent any identity overlap across training, validation, and testing sets. Specifically, the IG-AD100 dataset was divided into 80\% training, 10\% validation, and 10\% in-device testing, balanced across the two age groups (4--9 and 10--16 years). All iCAM T10 samples were reserved for cross-sensor testing, and no subjects appeared in either sensor.

The same evaluation metrics used in the main experiments were applied here, including macro-averaged classification accuracy, per-class precision, recall, and F1-score. Regression performance was evaluated using MAE, RMSE, and accuracy within $\pm$1 and $\pm$2 years of the ground truth age. Model architectures, training configurations, data augmentations, and loss functions were held constant across experiments to ensure a consistent comparison.

\section{Results}
\label{sec:results}

\subsection{Eye-Based Model Performance}
We first report results using grayscale eye images as input. Table~\ref{tab:eye_overall_results} summarizes classification and regression performance across all six models. MobileNetV3-Large achieved the best overall performance, with the highest classification accuracy (83.82\%), lowest MAE (1.33), and highest $\pm$1-year age accuracy (84.56\%). EfficientNet-B3 matched MobileNetV3 in regression MAE but had a slightly lower classification F1-score. DenseNet-121 and ResNet-50 performed competitively, though both exhibited increased error in regression.

\begin{table*}[h]
\centering
\caption{Overall classification and regression performance using iris images.}
\label{tab:iris_results}
\begin{tabular}{|c|c|c|c|c|}
\hline
\textbf{Model} & \textbf{Accuracy (\%)} & \textbf{MAE} & \textbf{RMSE} & \textbf{F1-Score (Avg)} \\
\hline
MobileViT-S        & \textbf{72.45} & \textbf{2.03} & \textbf{2.56} & \textbf{0.72} \\
\hline
EfficientNet-B3    & 72.38 & 2.29 & 8.55 & 0.75 \\
\hline
MobileNetV3-Large  & 71.00 & 2.32 & 2.94 & 0.72 \\
\hline
ResNet-50          & 68.56 & 2.26 & 2.81 & 0.68 \\
\hline
DenseNet-121       & 68.98 & 2.29 & 2.88 & 0.67 \\
\hline
ConvNeXt-Tiny      & 56.07 & 2.46 & 3.07 & 0.52 \\
\hline
\end{tabular}
\end{table*}

MobileViT-S demonstrated comparable classification performance but exhibited greater variability in regression metrics across age groups. ConvNeXt-Tiny underperformed significantly across all metrics, failing to learn age-relevant features for this task.

Per-class results, shown in Table~\ref{tab:eye_classwise_metrics}, indicate that MobileNetV3-Large and EfficientNet-B3 maintained balanced precision and recall across both age groups. Both models achieved F1-scores above 0.80 in Class 0 (ages 4--9) and Class 1 (ages 10--16). DenseNet-121 and ResNet-50 exhibited a slight bias toward younger subjects, with lower performance in older age groups. ConvNeXt-Tiny failed to predict Class 0 entirely, defaulting to majority class outputs, which aligns with its poor overall performance.

\begin{figure*}[!ht]
\centering
\includegraphics[width=\linewidth]{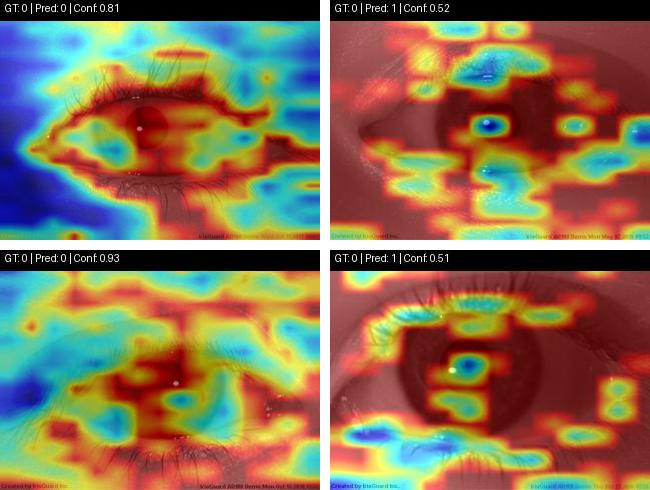}
\caption{Grad-CAM visualizations from the eye-based MobileNetV3 model. Each image is annotated with the ground truth class (GT), predicted class (Pred), and classification confidence (Conf). The left column shows correctly classified examples with focused attention on periocular regions such as eyelid folds, scleral boundaries, and lash contours features known to evolve with age. The right column shows misclassified samples, where the model attention is scattered or misaligned, often falling on noninformative regions such as background skin or upper brow. These failure cases highlight the model’s sensitivity to image ambiguity and its reliance on high-quality anatomical cues for accurate prediction.}

\label{fig:eye_gradcam}
\end{figure*}

To further interpret model behavior, we visualized Grad-CAM heatmaps generated from the classification stream of the MobileNetV3 model. As shown in Figure~\ref{fig:eye_gradcam}, the model consistently attends to the eyelid folds, periocular skin, and scleral boundaries, the regions known to undergo age related morphological changes. In contrast to the iris-based model, these activations are spatially diverse and context rich, supporting the hypothesis that periocular features encode strong age discriminative cues during childhood and adolescence. Notably, attention patterns were especially prominent near the upper eyelid crease and lower tarsal plate, suggesting reliance on skin tension and eye shape dynamics that evolve with age.

\subsection{Iris-Based Model Performance}
We also evaluated all models using normalized iris images, with corresponding binary masks provided as a second input channel. Table~\ref{tab:iris_results} summarizes the overall classification and regression performance for all models trained on iris data.

Across all architectures, iris-based models consistently underperformed relative  to their eye-based counterparts. MobileViT-S achieved the highest classification accuracy (72.45\%) and the lowest regression error (MAE = 2.03). EfficientNet-B3 and MobileNetV3-Large also surpassed 70\% accuracy but showed higher regression variance. ResNet-50 and DenseNet-121 delivered moderate results, while ConvNeXt-Tiny remained the lowest performer with an accuracy of 56.07\% and the highest regression error.

The lower performance of iris models is primarily attributed to the limited spatial resolution and the absence of surrounding periocular features, which often carry age-relevant cues such as skin texture, eyelid structure, and scleral contrast. These limitations were particularly pronounced among younger subjects, where age-related changes are more subtle and context-dependent. As a result, regression error remained consistently higher, and classification precision was reduced across all models.

Due to the reduced predictive reliability of iris models, we do not conduct further breakdowns such as age-bin analysis or confidence-based evaluation for this modality. All subsequent analyses focus on models trained using eye images.

\begin{figure*}[!ht]
\centering
\includegraphics[width=\linewidth]{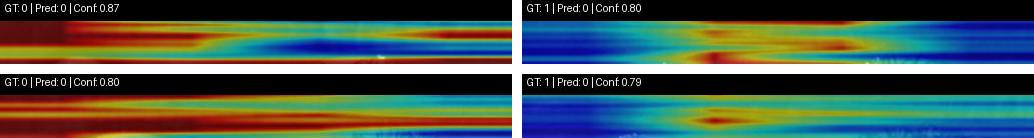}
\caption{Grad-CAM visualizations from the iris-based MobileNetV3 model using normalized iris images. The left column displays correctly classified samples, while the right column shows misclassified cases. Each image includes the ground truth class (GT), predicted class (Pred), and classification confidence (Conf). In correct predictions, attention maps reveal consistent horizontal banding aligned with the normalized iris texture, with moderate focus on stroma and crypt patterns. In contrast, incorrect predictions show weak or fragmented attention, often failing to capture distinctive anatomical cues. These patterns suggest that while iris texture may contain some age discriminative cues, they are insufficiently localized or consistent for high confidence classification in pediatric populations.}
\label{fig:iris_gradcam}
\end{figure*}

To better understand the spatial behavior of iris-based models, we visualized Grad-CAM attention maps derived from the MobileNetV3-Large classification stream. As shown in Figure~\ref{fig:iris_gradcam}, model focus was primarily constrained to the inner iris and pupil border, with weak or diffuse activation in peripheral regions. This spatial confinement likely limits the model's ability to capture age discriminative signals, especially in pediatric populations where developmental cues may lie outside the normalized iris annulus. While these attention maps do show mild emphasis on iris crypts and stroma, they further reinforce the hypothesis that age relevant texture changes, though present, are insufficient in isolation to support robust classification.

\subsection{Modality Comparison: Eye vs. Iris}

To assess the relative effectiveness of different input modalities, we compared the performance of each model trained on full-eye images versus normalized iris images. Table~\ref{tab:modality_comparison} presents a head-to-head comparison across classification and regression metrics.

Across all architectures, eye-based models consistently outperformed their iris-based counterparts. For example, MobileNetV3-Large improved from 71.00\% to 83.82\% in classification accuracy and reduced its MAE from 2.32 to 1.33 when trained on eye inputs. Similar gains were observed for EfficientNet-B3 (72.38\% to 80.58\% accuracy) and DenseNet-121 (68.98\% to 80.22\%).

This performance gap is primarily attributed to the richer spatial and contextual information present in eye images. These include periocular skin, eyelids, eyelashes, and scleral boundaries, features that often carry age-discriminative cues. In contrast, iris inputs are tightly normalized to a $256\times32$ strip, which restricts high-frequency texture information and omits surrounding facial context. Such limitations were particularly impactful in younger subjects, where subtle developmental signals are harder to capture.

These findings reinforce the superior predictive utility of periocular imagery and justify our decision to focus all subsequent analyses, including age-wise breakdowns, cross-sensor evaluations, and deployment benchmarking, exclusively on models trained with eye images.

\begin{table}[ht]
\centering
\caption{Performance comparison of each model across eye and iris inputs}
\begin{tabular}{|l|c|c|c|c|c|}
\hline
\multirow{2}{*}{\textbf{Model}} & \multicolumn{2}{c}{\textbf{Accuracy (\%)}} & \multicolumn{2}{c}{\textbf{MAE}} & \textbf{F1 (Avg)} \\
                                & Eye        & Iris        & Eye   & Iris  &        \\
\hline
\textbf{MobileNetV3-Large} & \textbf{83.82} & 71.00       & \textbf{1.33} & 2.32 & \textbf{0.83} \\
\hline
EfficientNet-B3            & 80.58         & 72.38       & 1.33         & 2.29 & 0.80 \\
\hline
MobileViT-S                & 77.12         & 72.45       & 1.55         & 2.03 & 0.74 \\
\hline
ResNet-50                  & 77.12         & 68.56       & 1.62         & 2.26 & 0.76 \\
\hline
DenseNet-121               & 80.22         & 68.98       & 1.54         & 2.29 & 0.77 \\
\hline
ConvNeXt-Tiny              & 45.23         & 56.07       & 2.53         & 2.46 & 0.00 \\
\hline
\end{tabular}
\label{tab:modality_comparison}
\end{table}

\subsection{Age-Wise Performance Analysis}

To evaluate model behavior across developmental stages, we analyzed regression performance within four age bins: 4--6, 7--9, 10--12, and 13--16 years. Table~\ref{tab:age_group_regression} reports the MAE, RMSE, and the percentage of predictions within $\pm$1 and $\pm$2 years of the ground truth for each model and age group.

\begin{figure*}[h]
    \centering
    \includegraphics[width=\linewidth]{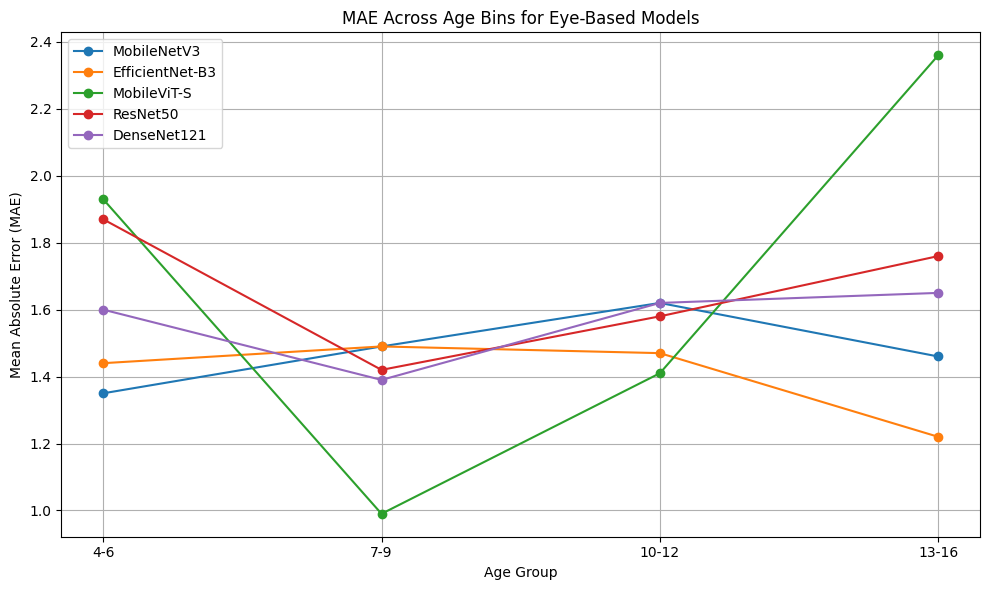}
    \caption{MAE across age bins for each eye-based model. MobileNetV3 and EfficientNet-B3 show stable performance, while MobileViT-S shows high variance.}
    \label{fig:mae_age_bins}
\end{figure*}

Figure~\ref{fig:mae_age_bins} presents a line plot of MAE across age bins. MobileNetV3-Large and EfficientNet-B3 demonstrate stable and consistent performance across all age groups, while MobileViT-S shows sharp variability, particularly excelling in the 7--9 group but degrading in the 13--16 group. ResNet-50 and DenseNet-121 maintain reasonable performance in the mid childhood range (7--12) but are less reliable in younger and older cohorts.

\begin{figure*}[h]
    \centering
    \includegraphics[width=\linewidth]{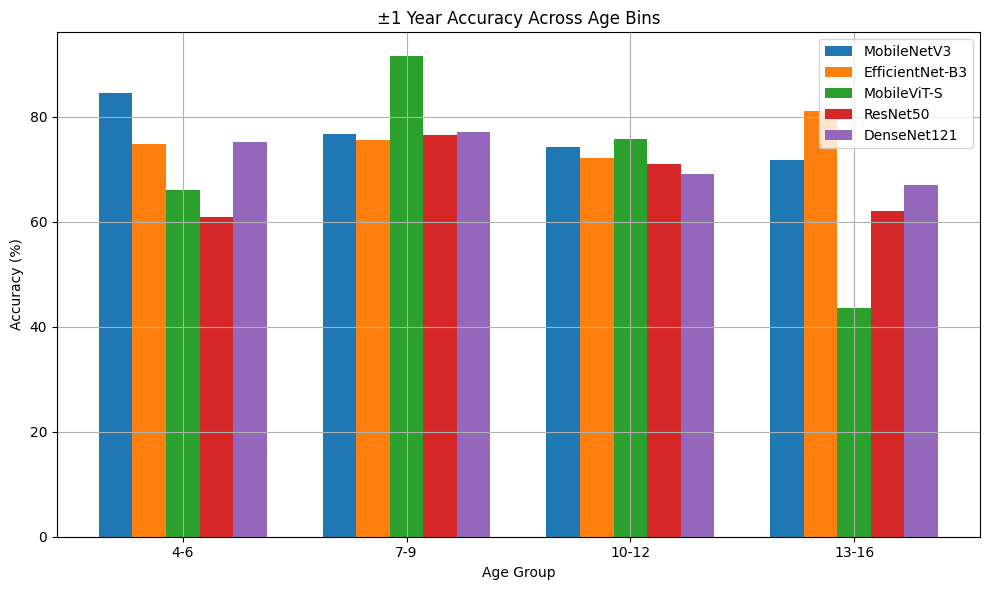}
    \caption{$\pm$1 year accuracy across age bins. EfficientNet-B3 excels in the 13--16 group, while MobileViT-S peaks in the 7--9 range but declines sharply afterward.}
    \label{fig:acc1_age_bins}
\end{figure*}

Figure~\ref{fig:acc1_age_bins} complements this analysis by illustrating $\pm$1 year accuracy across the same age bins. MobileNetV3 leads in early childhood (4--6), while EfficientNet-B3 achieves the best performance in adolescence (13--16). MobileViT-S displays the most fluctuation, peaking in the 7--9 bin and falling off sharply afterward.

Overall, results indicate that age estimation is more challenging in the youngest (4--6) and oldest (13--16) groups, while the mid-childhood range (7--12) is more predictable. MobileNetV3 and EfficientNet-B3 demonstrated the most robust cross-bin performance, supporting their suitability for deployment and downstream analysis.

\begin{table*}[ht]
\centering
\caption{Grouped Regression Performance Across Age Bins Using Eye Images}
\begin{tabular}{|l|l|c|c|c|c|}
\hline
\textbf{Model} & \textbf{Age Group} & \textbf{MAE} & \textbf{RMSE} & \textbf{Accuracy} (±1 yr (\%) & \textbf{Accuracy} (±2 yr) (\%) \\
\hline
\multirow{4}{*}{\textbf{MobileNetV3-Large}} 
    & 4–6  & 1.35 & 2.05 & \textbf{84.56} & 90.44 \\
    & 7–9  & 1.49 & 2.17 & 76.72 & 86.21 \\
    & 10–12 & 1.62 & 2.03 & 74.12 & 86.47 \\
    & 13–16 & 1.46 & 1.96 & 71.74 & 90.22 \\
\hline
\multirow{4}{*}{\textbf{EfficientNet-B3}} 
    & 4–6  & 1.44 & 2.15 & 74.71 & 88.81 \\
    & 7–9  & 1.49 & 1.96 & 75.55 & 88.81 \\
    & 10–12 & 1.47 & 1.82 & 72.19 & \textbf{91.54} \\
    & 13–16 & \textbf{1.22} & \textbf{1.55} & \textbf{81.11} & \textbf{95.28} \\
\hline
\multirow{4}{*}{\textbf{MobileViT-S}} 
    & 4–6  & 1.93 & 2.36 & 65.98 & 81.61 \\
    & 7–9  & \textbf{0.99} & \textbf{1.40} & \textbf{91.46} & 95.29 \\
    & 10–12 & 1.41 & 1.75 & 75.65 & 90.67 \\
    & 13–16 & 2.36 & 2.73 & 43.61 & 68.89 \\
\hline
\multirow{4}{*}{\textbf{ResNet-50}} 
    & 4–6  & 1.87 & 2.43 & 60.92 & 81.15 \\
    & 7–9  & 1.42 & 1.88 & 76.58 & 88.07 \\
    & 10–12 & 1.58 & 2.00 & 70.98 & 83.94 \\
    & 13–16 & 1.76 & 2.15 & 61.94 & 85.28 \\
\hline
\multirow{4}{*}{\textbf{DenseNet-121}} 
    & 4–6  & 1.60 & 2.25 & 75.17 & 85.52 \\
    & 7–9  & 1.39 & 1.87 & 77.03 & 89.54 \\
    & 10–12 & 1.62 & 2.04 & 69.08 & 85.15 \\
    & 13–16 & 1.65 & 2.01 & 66.94 & 87.78 \\
\hline
\end{tabular}
\label{tab:age_group_regression}
\end{table*}

\subsection{Confidence vs. Age Trends}

\begin{figure*}[h]
    \centering
    \includegraphics[width=0.85\linewidth]{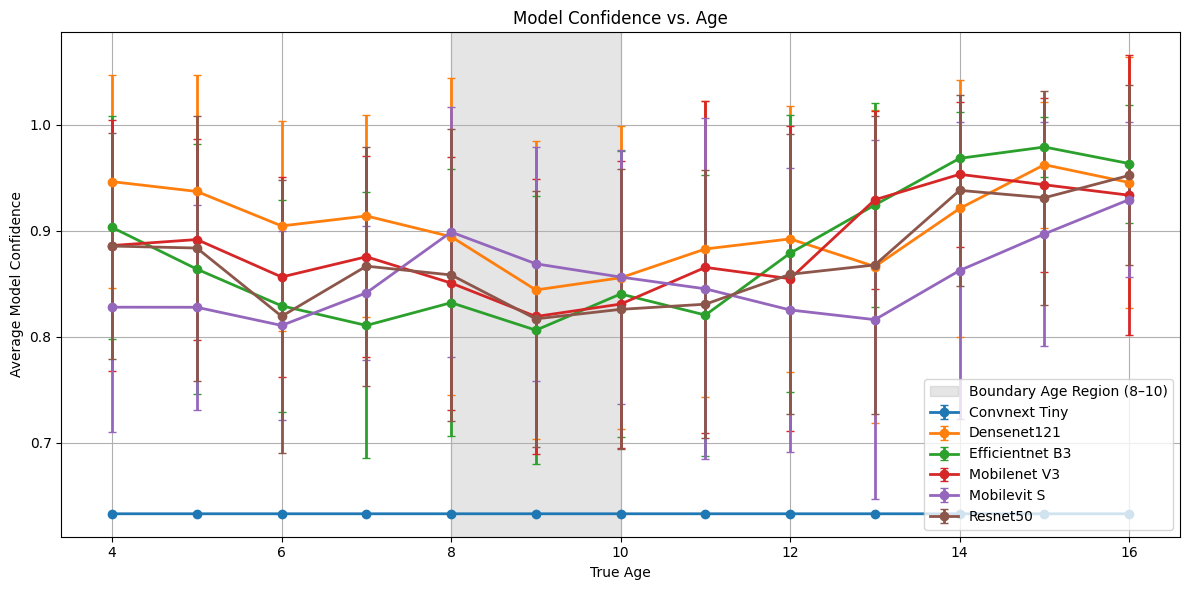}
    \caption{Average model confidence vs. true age using eye inputs. A consistent dip is observed around the 8--10-year boundary (shaded region), where inter-class distinctions are less defined.}
    \label{fig:confidence_age_curve}
\end{figure*}

To better understand how model certainty evolves across the pediatric age spectrum, we analyzed classification confidence as a function of actual subject age. Figure~\ref{fig:confidence_age_curve} shows the average softmax confidence for the predicted class, grouped by age (4 to 16 years), using eye-based inputs.

All models exhibited a consistent U-shaped trend: confidence scores were highest in early childhood (ages 4--6) and adolescence (13--16), but dipped notably in the transitional period between 8--10 years. This region is highlighted as a shaded band in Figure~\ref{fig:confidence_age_curve} represents the decision boundary between the two classification groups (4--9 vs. 10--16). Here, subtle visual changes make age prediction less specific and increase inter-class confusion.

MobileNetV3 and EfficientNet-B3 maintained the highest confidence across all ages and exhibited the shallowest dips in the boundary zone, indicating a stronger separation between learned features. MobileViT-S showed marked volatility between ages 9--12, aligning with its variable predictive performance. Meanwhile, ConvNeXt-Tiny exhibited flat, low confidence throughout the age spectrum, consistent with its overall weak learning capacity.

These trends underscore the importance of training strategies such as label smoothing and soft target boundaries, particularly near class transitions. Additionally, confidence scores offer a practical basis for uncertainty-aware logic. Systems can be designed to defer decisions or request secondary validation when operating in known ambiguity zones, such as the 8--10 age range.

\subsection{Cross-Sensor Generalization}

To evaluate model robustness under real-world deployment conditions, we conducted a cross-sensor experiment. Each model was trained exclusively on images from the IG-AD100 sensor and tested on two identity-exclusive sets: one from the same IG-AD100 device and another from the unseen iCAM T10 sensor. The same subject-level splitting protocol was used throughout.

Tables~\ref{tab:cross_sensor_igad100} and~\ref{tab:cross_sensor_icam} report the classification and regression results across both domains. As expected, all models experienced a performance drop when evaluated on iCAM T10 data. However, the degree of degradation varied substantially across architectures.

MobileNetV3-Large remained the most resilient, with classification accuracy improving slightly on the cross-sensor set (79.10\%) and MAE increasing modestly from 1.49 to 2.09. It also maintained strong $\pm$2-year accuracy (90.22\%) on the unseen sensor. EfficientNet-B3 delivered the highest cross-sensor performance overall, with 85.86\% accuracy, 1.37 MAE, and 96.98\% $\pm$2-year accuracy, indicating excellent generalization to sensor variability.

MobileViT-S showed competitive results on both sensors but exhibited increased error in the adolescent age group during cross-sensor evaluation. This may indicate that attention-based models are more sensitive to subtle distributional shifts caused by hardware differences. In contrast, ConvNeXt-Tiny, ResNet-50, and DenseNet-121 exhibited greater instability, with sharp declines in accuracy and elevated regression errors, particularly in younger subjects. ConvNeXt-Tiny failed to preserve meaningful class separation across devices, collapsing on both classification and regression tasks.

Grouped analysis further confirmed that MobileNetV3 and EfficientNet-B3 sustained strong $\pm$1-year and $\pm$2-year accuracy across all age groups, even under hardware variation. These findings underscore the importance of selecting architectures that balance representational power with robustness when deploying in heterogeneous sensor environments.

\begin{table*}[h]
\centering
\caption{Cross-sensor evaluation, models trained on IG-AD100 and tested on IG-AD100 (same sensor).}
\label{tab:cross_sensor_igad100}
\begin{tabular}{|l|c|c|c|c|c|c|}
\hline
\textbf{Model} & \textbf{Accuracy (\%)} & \textbf{MAE} & \textbf{RMSE} & \textbf{F1 (Avg)} & \textbf{$\pm$1yr (\%)} & \textbf{$\pm$2yr (\%)} \\
\hline
MobileNetV3-Large  & 78.25 & 1.49 & 1.87 & 0.75 & 68.23 & 89.30 \\
\hline
EfficientNet-B3    & 79.98 & 1.34 & 1.76 & 0.77 & 73.58 & 84.62 \\
\hline
MobileViT-S        & 80.17 & 1.34 & 1.80 & 0.77 & 72.24 & 82.94 \\
\hline
ConvNeXt-Tiny      & 55.54 & 2.46 & 2.98 & 0.57 & 0.00  & 1.67  \\
\hline
ResNet-50          & 72.18 & 2.34 & 2.86 & 0.73 & 25.75 & 53.85 \\
\hline
DenseNet-121       & 75.89 & 1.91 & 2.35 & 0.75 & 48.83 & 68.56 \\
\hline
\end{tabular}
\end{table*}

\begin{table*}[h]
\centering
\caption{Cross-sensor evaluation, models trained on IG-AD100 and tested on iCAM T10 (unseen sensor).}
\label{tab:cross_sensor_icam}
\begin{tabular}{|l|c|c|c|c|c|c|}
\hline
\textbf{Model} & \textbf{Accuracy (\%)} & \textbf{MAE} & \textbf{RMSE} & \textbf{F1 (Avg)} & \textbf{$\pm$1yr (\%)} & \textbf{$\pm$2yr (\%)} \\
\hline
MobileNetV3-Large  & 79.10 & 2.09 & 2.56 & 0.78 & 71.74 & 90.22 \\
\hline
EfficientNet-B3    & 85.86 & 1.37 & 1.75 & 0.87 & 87.84 & 96.98 \\
\hline
MobileViT-S        & 80.96 & 1.72 & 2.16 & 0.81 & 74.25 & 91.42 \\
\hline
ConvNeXt-Tiny      & 51.71 & 3.00 & 3.63 & 0.59 & 27.99 & 51.71 \\
\hline
ResNet-50          & 70.00 & 2.34 & 2.96 & 0.75 & 66.79 & 82.09 \\
\hline
DenseNet-121       & 72.36 & 2.03 & 2.54 & 0.76 & 66.42 & 87.31 \\
\hline
\end{tabular}
\end{table*}

\subsection{Deployment Benchmarking}
To assess real-world applicability, we evaluated the inference latency and model footprint across two deployment contexts: (1) high-performance and embedded hardware (RTX TITAN and Jetson AGX Orin), and (2) immersive standalone VR environments (Oculus Quest 2), where only MobileNetV3 was deployed.

Table~\ref{tab:inference_time_model_size} summarizes model sizes and average inference times measured on both RTX TITAN and Jetson AGX Orin. Among all models, MobileNetV3 exhibited the most efficient deployment profile, with the smallest model size of 17 MB and the fastest inference speed of just 0.07 ms on RTX TITAN and 3.29 ms on Jetson. MobileViT-S and EfficientNet-B3 also achieved sub-15 ms latency on Jetson, although with slightly larger memory footprints. In contrast, ResNet-50 and DenseNet-121 incurred higher latency due to their deeper and wider structures, making them less suitable for constrained devices.

\begin{table}[ht]
\centering
\caption{Deployment Benchmarking: Model Size and Inference Latency Across Devices}
\label{tab:deployment_benchmark}
\begin{tabularx}{\linewidth}{|l|X|X|X|}
\hline
\textbf{Model} & \textbf{Size (MB)} & \textbf{RTX TITAN (ms)} & \textbf{Jetson Orin (ms)} \\
\hline
\textbf{MobileNetV3-Large} & \textbf{17.0} & \textbf{0.07} & \textbf{3.29} \\
MobileViT-S                & 19.9          & 0.08          & 9.23 \\
\hline
EfficientNet-B3            & 43.4          & 0.10          & 12.12 \\
\hline
DenseNet-121               & 28.4          & 0.12          & 15.87 \\
\hline
ResNet-50                  & 94.4          & 0.15          & 12.65 \\
\hline
\end{tabularx}
\end{table}

For VR deployment, MobileNetV3 was selected due to its favorable trade-off between runtime efficiency and predictive accuracy. The model was trained in FP32, fine-tuned with PyTorch’s Automatic Mixed Precision (AMP), and exported to ONNX. It was subsequently converted to TensorFlow Lite (TFLite) with FP16 quantization for reduced memory usage and optimized inference.

The quantized TFLite model was integrated into the Unity Engine (2022.3 LTS) using Unity’s Barracuda inference framework and deployed on the Oculus Quest 2 headset, powered by the Qualcomm Snapdragon XR2. Using the Oculus Performance Toolkit, the average inference latency was measured under a sustained load at 10 ms per image, well within the real-time constraints for VR use cases.

Among all tested models, MobileNetV3 demonstrated the most substantial deployment potential, balancing compact size, competitive accuracy, and sub-20 ms latency across desktop, edge, and immersive platforms. These results support its suitability for real-time, low-power, and on-device pediatric age estimation.

\section{Discussion}
\label{sec:discussion}

This study presents one of the first in-depth investigations into pediatric age estimation using ocular biometrics, leveraging a rich, longitudinal dataset collected over eight years. By analyzing normalized iris patterns alongside wider periocular images, we explored how these two modalities perform across age groups, imaging devices, and real-world deployment scenarios. The aim wasn’t just accuracy but it was about building systems that could actually work for children, across time, hardware, and developmental stages.

What stood out consistently was the advantage of periocular images. Compared to iris inputs, models trained on eye images performed better across the board. That makes sense: the periocular region brings in more than just texture like it includes eyelid shapes, scleral boundaries, and subtle skin cues. These are features that shift with age, especially in younger children. Iris textures alone, which are often reduced to narrow 1D strips in normalized form, seem to miss that nuanced variability. This resulted in lower accuracy, particularly for children under 10, where facial and ocular changes are subtle but important.

Among all models we evaluated, MobileNetV3 consistently struck the best balance. Despite its compact size, it delivered strong classification and regression results, even when tested on data from different sensors. Its efficient design allowed sub-20 millisecond inference across all platforms, from desktop GPUs to embedded systems like Jetson AGX Orin, and even standalone VR headsets. That kind of speed, combined with reliable accuracy, makes it a strong candidate for practical deployment in safety critical or resource-constrained environments, like VR-based age verification systems.

But there’s a caveat. Our runtime benchmarks assume the availability of NIR ocular images. Most consumer VR devices, such as the Oculus Quest 2, only have RGB cameras for tracking. For real-world deployment, this means adding a lightweight NIR module to the headset, either built-in or via a peripheral. Interestingly, newer devices like the Meta Quest Pro and Apple Vision Pro already include eye-tracking hardware. With proper tuning to NIR wavelengths, these could potentially support our system with minimal hardware changes.

Our age-wise confidence analysis revealed something curious. There’s a noticeable dip in model certainty between the 4–9 and 10–16 age brackets, mainly around ages 8 to 10. These transitional years seem to blur the visual cues that separate younger and older children. The result is increased ambiguity and overlapping features, suggesting that hard classification boundaries may not be ideal. Instead, we may need uncertainty-aware models that defer decisions or bring in additional modalities when confidence drops.

Cross-sensor evaluation also provided valuable insights. While all models saw some performance drop on unseen devices, MobileNetV3 and EfficientNet-B3 handled the domain shift gracefully. Others, like ResNet-50 and MobileViT-S, were more sensitive, particularly when classifying younger children. This points to a need for models that are either inherently sensor-agnostic or trained with domain adaptation strategies.

Performance varied by age group as well. Predictably, the youngest (4–6) and oldest (13–16) groups were the most challenging. Early childhood is a period of rapid change, leading to high within group variability. Adolescents, on the other hand, show more inter-subject differences, making generalization harder. In contrast, the 7–12 age range yielded more stable predictions, perhaps due to more uniform growth patterns during those years. This suggests that longitudinal modeling or temporal attention mechanisms could help stabilize predictions, especially at the edges of the developmental curve.

It’s also worth noting the limitations of our setup. All images were captured in grayscale NIR under semi-controlled conditions. While this allowed us to standardize acquisition and ensure safety, it doesn’t fully represent in-the-wild settings. Moreover, age labels were based on birth year metadata, which introduces some granularity issues, especially for regression tasks. We mitigated these by focusing on age classification bins and incorporating interpretability tools like Grad-CAM.

These visualizations proved illuminating. Periocular models consistently focused on meaningful areas like eyelids, skin folds, and scleral boundaries. Iris-based models, on the other hand, often locked onto limited regions like horizontal stroma bands or the pupil edge in misclassified cases. The lack of spatial diversity in their attention maps aligned with their lower performance. Interestingly, even in failure cases, periocular models showed scattered but contextually relevant focus, which could be useful for identifying ambiguous samples or triggering fallback mechanisms.

A significant finding was the moderate yet reliable performance of iris-based models. While they did not match their periocular counterparts, they were not completely off the mark either. This leads us to explore the possibility that age-linked changes in iris stroma, pigmentation, or dilation behavior may be driving some of this signal. Prior work has suggested that certain fine-grained features within the iris, such as stroma density and pigmentation patterns, may undergo gradual evolution over the human lifespan~\cite{fairhurst2011analysis, rajput2019deep}. Although most longitudinal studies have focused on adult populations, it is plausible that similar developmental cues are subtly present even in early childhood and adolescence, contributing to the moderate performance of iris-based age models observed in our study. Traditionally, the iris is treated as a stable identity trait, but our results suggest there may be subtle age-related features that CNNs can pick up on, especially in children. This doesn’t challenge the foundational stability of iris biometrics for identity verification. Instead, it invites further inquiry: can seemingly invariant biometric traits encode auxiliary information like age?

Anticipating future developments, several directions seem promising. First, a deeper dive into the biological underpinnings of iris texture changes during childhood might clarify whether we’re observing real developmental signals or merely data driven artifacts. Cross-spectral studies between visible and NIR imaging, combined with interpretability tools and anatomical insights, could explain more here.

Second, fusing iris and periocular features or even tapping into other modalities might yield better age predictions, particularly in edge cases. There’s also a case for integrating soft decision mechanisms, bias auditing across demographics, and semi-supervised learning techniques to reduce the dependency on large labeled pediatric datasets.

Despite its scope, this work lays an important foundation. We offer a scalable, interpretable framework for real-time pediatric age estimation, one that’s practical, hardware-aware, and deeply rooted in eight years of data. In contexts where age matters but formal IDs are unavailable like VR classrooms, educational games, or pediatric safety systems, such solutions could be transformative.

\section{Conclusion}
\label{sec:conclusion}
This study provides a comprehensive benchmark for pediatric age estimation using ocular biometrics, based on a unique eight-year longitudinal dataset of NIR images. By comparing normalized iris and periocular inputs across multiple deep learning architectures, age groups, sensors, and deployment contexts, we demonstrate the clear advantages of eye-based models in capturing age discriminative cues.

Among all models, MobileNetV3 achieved the best trade-off between accuracy and latency, demonstrating real-time performance even on resource constrained VR hardware. Confidence and error trend analyses further highlighted model uncertainty near developmental boundaries, offering design insights for robust, real-world deployment in child-focused applications.

\section*{Acknowledgment}
This section will be completed following acceptance.

\bibliographystyle{IEEEtran}\
\bibliography{bibliography}

\begin{thebibliography}{10}
\providecommand{\url}[1]{#1}
\csname url@samestyle\endcsname
\providecommand{\newblock}{\relax}
\providecommand{\bibinfo}[2]{#2}
\providecommand{\BIBentrySTDinterwordspacing}{\spaceskip=0pt\relax}
\providecommand{\BIBentryALTinterwordstretchfactor}{4}
\providecommand{\BIBentryALTinterwordspacing}{\spaceskip=\fontdimen2\font plus
\BIBentryALTinterwordstretchfactor\fontdimen3\font minus \fontdimen4\font\relax}
\providecommand{\BIBforeignlanguage}[2]{{%
\expandafter\ifx\csname l@#1\endcsname\relax
\typeout{** WARNING: IEEEtran.bst: No hyphenation pattern has been}%
\typeout{** loaded for the language `#1'. Using the pattern for}%
\typeout{** the default language instead.}%
\else
\language=\csname l@#1\endcsname
\fi
#2}}
\providecommand{\BIBdecl}{\relax}
\BIBdecl

\bibitem{Ruth}
M.~S. M.~A. Ruth, L.~Gita, A.~Kurniawan, H.~Utomo \emph{et~al.}, ``Age estimation with smartphone: Is it reliable for forensics identification? estimasi usia dengan smartphone: Apakah dapat diandalkan dalam identifkasi forensik?'' \emph{Dentika: Dental Journal}, vol.~23, no.~2, pp. 34--38, 2020.

\bibitem{fiani2024exploring}
C.~Fiani, P.~Saeghe, M.~McGill, and M.~Khamis, ``Exploring the perspectives of social vr-aware non-parent adults and parents on children's use of social virtual reality,'' \emph{Proceedings of the ACM on Human-Computer Interaction}, vol.~8, no. CSCW1, pp. 1--25, 2024.

\bibitem{Liao}
H.~Liao, Y.~Yan, W.~Dai, and P.~Fan, ``Age estimation of face images based on cnn and divide-and-rule strategy,'' \emph{Mathematical Problems in Engineering}, vol. 2018, no.~1, p. 1712686, 2018.

\bibitem{Singh}
S.~Singh, K.~Bahmani, and S.~Schuckers, ``Longitudinal evaluation of child face recognition and the impact of underlying age,'' in \emph{2024 IEEE International Joint Conference on Biometrics (IJCB)}.\hskip 1em plus 0.5em minus 0.4em\relax IEEE, 2024, pp. 1--9.

\bibitem{bisogni2022periocular}
C.~Bisogni, L.~Cascone, and F.~Narducci, ``Periocular data fusion for age and gender classification,'' \emph{Journal of Imaging}, vol.~8, no.~11, p. 307, 2022.

\bibitem{fairhurst2011analysis}
M.~Fairhurst and M.~Erbilek, ``Analysis of physical ageing effects in iris biometrics,'' \emph{IET Computer Vision}, vol.~5, no.~6, pp. 358--366, 2011.

\bibitem{Mehrotra}
H.~Mehrotra, M.~Vatsa, R.~Singh, and B.~Majhi, ``Does iris change over time?'' \emph{PloS one}, vol.~8, no.~11, p. e78333, 2013.

\bibitem{Erbilek}
M.~Erbilek, M.~Fairhurst, and M.~C. D.~C. Abreu, ``Age prediction from iris biometrics,'' in \emph{5th International Conference on Imaging for Crime Detection and Prevention (ICDP 2013)}.\hskip 1em plus 0.5em minus 0.4em\relax IET, 2013, pp. 1--07.

\bibitem{odion2022age}
P.~Odion, M.~Musa, and S.~Shuaibu, ``Age prediction from sclera images using deep learning,'' \emph{Journal of the Nigerian Society of Physical Sciences}, pp. 787--787, 2022.

\bibitem{Hashemi}
H.~Hashemi, R.~Pakzad, M.~Khabazkhoob, A.~Yekta, M.~H. Emamian, and A.~Fotouhi, ``Ocular biometrics as a function of age, gender, height, weight, and its association with spherical equivalent in children,'' \emph{European Journal of Ophthalmology}, vol.~31, no.~2, pp. 688--697, 2021.

\bibitem{Zengin}
M.~Zengin, E.~Karahan, S.~Yilmaz, E.~Cinar, I.~Tuncer, and C.~Kucukerdonmez, ``Association of choroidal thickness with eye growth: a cross-sectional study of individuals between 4 and 23 years,'' \emph{Eye}, vol.~28, no.~12, pp. 1482--1487, 2014.

\bibitem{Read}
S.~A. Read, D.~Alonso-Caneiro, and S.~J. Vincent, ``Longitudinal changes in macular retinal layer thickness in pediatric populations: Myopic vs non-myopic eyes,'' \emph{PLoS One}, vol.~12, no.~6, p. e0180462, 2017.

\bibitem{sgroi2013prediction}
A.~Sgroi, K.~W. Bowyer, and P.~J. Flynn, ``The prediction of old and young subjects from iris texture,'' in \emph{2013 International Conference on Biometrics (ICB)}.\hskip 1em plus 0.5em minus 0.4em\relax IEEE, 2013, pp. 1--5.

\bibitem{Alonso}
F.~Alonso-Fernandez, K.~Hernandez-Diaz, S.~Ramis, F.~J. Perales, and J.~Bigun, ``Facial masks and soft-biometrics: Leveraging face recognition cnns for age and gender prediction on mobile ocular images,'' \emph{IET Biometrics}, vol.~10, no.~5, pp. 562--580, 2021.

\bibitem{Gangwar}
A.~Gangwar and A.~Joshi, ``Deepirisnet: Deep iris representation with applications in iris recognition and cross-sensor iris recognition,'' in \emph{2016 IEEE international conference on image processing (ICIP)}.\hskip 1em plus 0.5em minus 0.4em\relax IEEE, 2016, pp. 2301--2305.

\bibitem{Sagnier}
C.~Sagnier, E.~Loup-Escande, D.~Lourdeaux, I.~Thouvenin, and G.~Vall{\'e}ry, ``User acceptance of virtual reality: an extended technology acceptance model,'' \emph{International Journal of Human--Computer Interaction}, vol.~36, no.~11, pp. 993--1007, 2020.

\bibitem{vrchat}
\BIBentryALTinterwordspacing
vrchat. Accessed: 24 March 2025. [Online]. Available: \url{https://wiki.vrchat.com/wiki/Age\_Verification}
\BIBentrySTDinterwordspacing

\bibitem{persona}
\BIBentryALTinterwordspacing
persona. Accessed: 24 March 2025. [Online]. Available: \url{https://withpersona.com/}
\BIBentrySTDinterwordspacing

\bibitem{das2021iris}
P.~Das, L.~Holsopple, D.~Rissacher, M.~Schuckers, and S.~Schuckers, ``Iris recognition performance in children: A longitudinal study,'' \emph{IEEE Transactions on Biometrics, Behavior, and Identity Science}, vol.~3, no.~1, pp. 138--151, 2021.

\bibitem{das2023longitudinal}
P.~Das, N.~G. Venkataswamy, L.~Holsopple, M.~H. Imtiaz, M.~Schuckers, and S.~Schuckers, ``Longitudinal performance of iris recognition in children: Time intervals up to six years,'' in \emph{2023 11th International Workshop on Biometrics and Forensics (IWBF)}.\hskip 1em plus 0.5em minus 0.4em\relax IEEE, 2023, pp. 1--6.

\bibitem{rajput2019deep}
M.~Rajput and G.~Sable, ``Deep learning based gender and age estimation from human iris,'' in \emph{Proceedings of the international conference on advances in electronics, electrical and computational intelligence (ICAEEC)}, 2019.

\bibitem{Hansen}
M.~H. Hansen, X.~Q. Li, M.~Larsen, E.~M. Olsen, A.~M. Skovgaard, L.~Kessel, and I.~C. Munch, ``Five-year change in choroidal thickness in relation to body development and axial eye elongation: the ccc2000 eye study,'' \emph{Investigative ophthalmology and visual science}, vol.~60, no.~12, pp. 3930--3936, 2019.

\bibitem{Hartstein}
L.~E. Hartstein, M.~K. LeBourgeois, M.~T. Durniak, and R.~P. Najjar, ``Differences in the pupillary responses to evening light between children and adolescents,'' \emph{Journal of Physiological Anthropology}, vol.~43, no.~1, p.~16, 2024.

\bibitem{meta}
\BIBentryALTinterwordspacing
meta. Accessed: 24 March 2025. [Online]. Available: \url{https://developers.meta.com/horizon}
\BIBentrySTDinterwordspacing

\bibitem{alhichri2021classification}
H.~Alhichri, A.~S. Alswayed, Y.~Bazi, N.~Ammour, and N.~A. Alajlan, ``Classification of remote sensing images using efficientnet-b3 cnn model with attention,'' \emph{IEEE access}, vol.~9, pp. 14\,078--14\,094, 2021.

\bibitem{koonce2021mobilenetv3}
B.~Koonce and B.~Koonce, ``Mobilenetv3,'' \emph{Convolutional Neural Networks with Swift for Tensorflow: Image Recognition and Dataset Categorization}, pp. 125--144, 2021.

\bibitem{koonce2021resnet}
B.~Koonce, ``Resnet 50,'' in \emph{Convolutional neural networks with swift for tensorflow: image recognition and dataset categorization}.\hskip 1em plus 0.5em minus 0.4em\relax Springer, 2021, pp. 63--72.

\bibitem{Huang}
G.~Huang, Z.~Liu, L.~Van Der~Maaten, and K.~Q. Weinberger, ``Densely connected convolutional networks,'' in \emph{Proceedings of the IEEE conference on computer vision and pattern recognition}, 2017, pp. 4700--4708.

\bibitem{liu2022convnet}
Z.~Liu, H.~Mao, C.-Y. Wu, C.~Feichtenhofer, T.~Darrell, and S.~Xie, ``A convnet for the 2020s,'' in \emph{Proceedings of the IEEE/CVF conference on computer vision and pattern recognition}, 2022, pp. 11\,976--11\,986.

\bibitem{mehta2021mobilevit}
S.~Mehta and M.~Rastegari, ``Mobilevit: light-weight, general-purpose, and mobile-friendly vision transformer,'' \emph{arXiv preprint arXiv:2110.02178}, 2021.

\bibitem{irisguard}
\BIBentryALTinterwordspacing
iris guard. Accessed: 24 March 2025. [Online]. Available: \url{https://www.irisguard.com}
\BIBentrySTDinterwordspacing

\bibitem{irisID}
\BIBentryALTinterwordspacing
iris ID. Accessed: 24 March 2025. [Online]. Available: \url{https://www.irisid.com/}
\BIBentrySTDinterwordspacing

\bibitem{othman2016osiris}
N.~Othman, B.~Dorizzi, and S.~Garcia-Salicetti, ``Osiris: An open source iris recognition software,'' \emph{Pattern recognition letters}, vol.~82, pp. 124--131, 2016.

\bibitem{daugman2009iris}
J.~Daugman, ``How iris recognition works,'' in \emph{The essential guide to image processing}.\hskip 1em plus 0.5em minus 0.4em\relax Elsevier, 2009, pp. 715--739.

\end{thebibliography}

\begin{IEEEbiography}[{\includegraphics[width=1in,height=1.25in,clip,keepaspectratio]{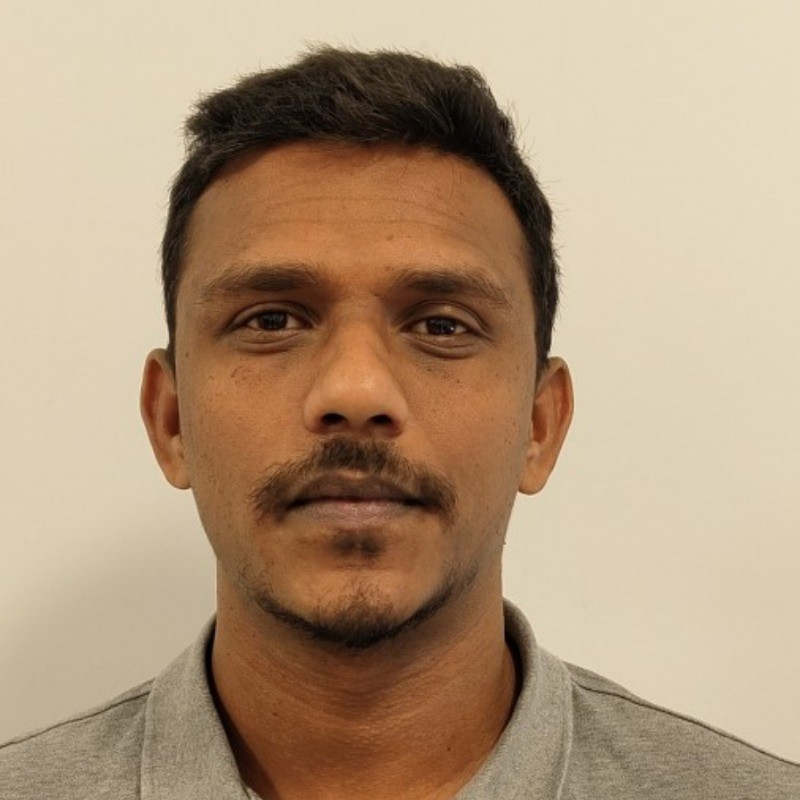}}]{Naveenkumar G. Venkataswamy} received a B.E. degree in Electrical and Communication Engineering from Visvesvaraya Technological University, Belagavi, India. He is currently pursuing a Ph.D. in Electrical and Computer Engineering at Clarkson University, with a research focus on biometrics, artificial intelligence, computer vision, and deep learning.
\end{IEEEbiography}

\begin{IEEEbiography}[{\includegraphics[width=1in,height=1.25in,clip,keepaspectratio]{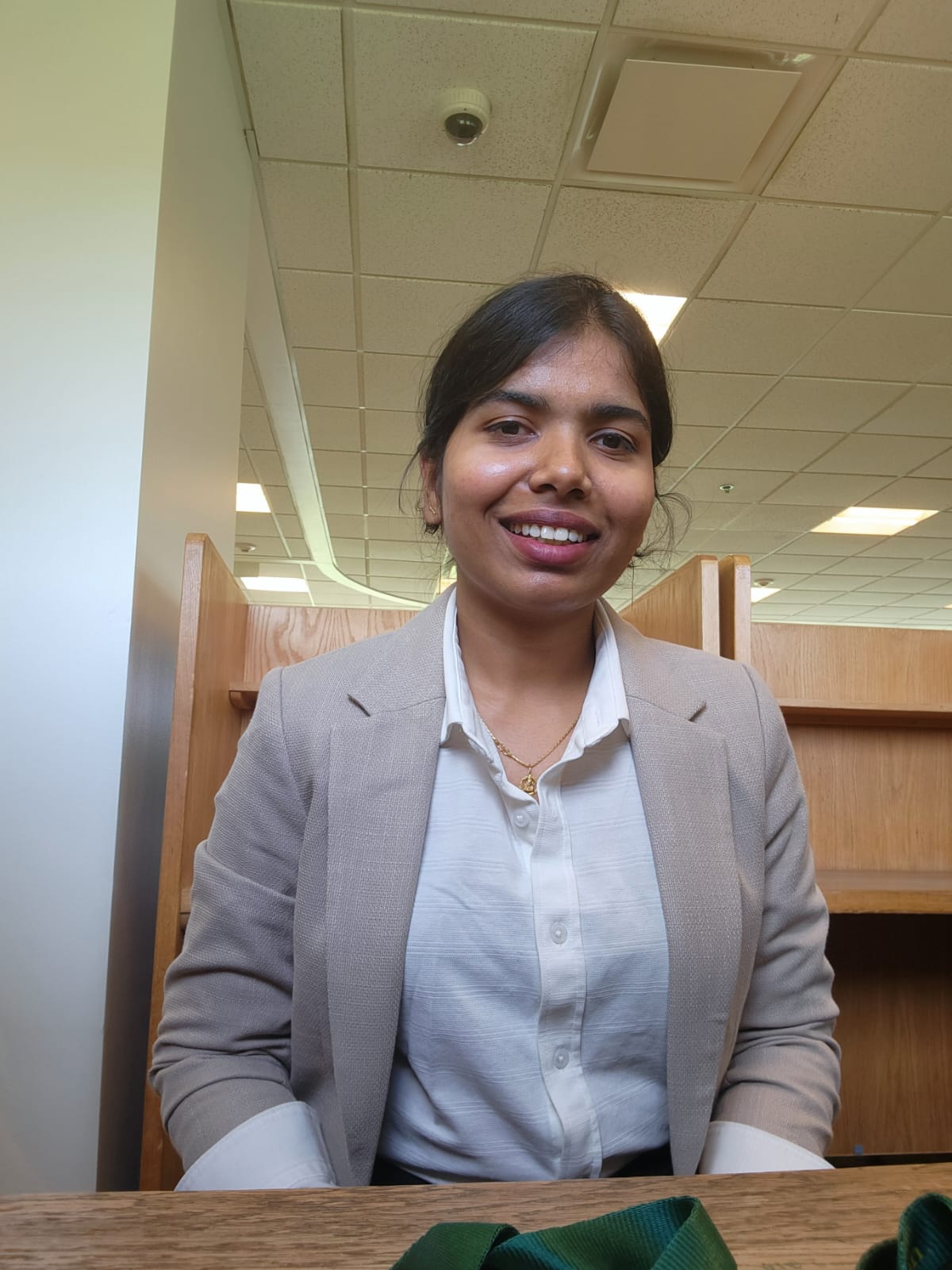}}]{Poorna Raavi} received a B. Tech degree in Computer Science from Acharya Nagarajuna University, Andhra Pradesh, India. She is currently pursuing a Master's degree in Computer Science at Clarkson University, with a research focus on synthetic voice, virtual reality development and deep learning.

\end{IEEEbiography}

\begin{IEEEbiography}[{\includegraphics[width=1in,height=1.25in,clip,keepaspectratio]{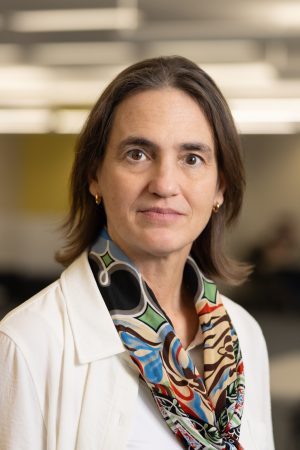}}]{Stephanie Schuckers} is the Bank of America Distinguished Professor in Computing \& Informatics at UNC Charlotte.  She is also the Visiting Research Professor in the Department of Electrical and Computer Engineering at Clarkson University and serves as the Director of the Center of Identification Technology Research (CITeR), a National Science Foundation Industry/University Cooperative Research Center. She received her doctoral degree in Electrical Engineering from The University of Michigan. Professor Schuckers research focuses on processing and interpreting signals which arise from the human body.  Her work is funded from various sources, including National Science Foundation, Department of Homeland Security, and private industry, among others.  She has started her own business, testified for US Congress, and has over 50 journal publications as well as over 100 other academic publications.  She was named IEEE Fellow in 2023, serves as a Board of Directors for the Biometrics Institute, and is President for the IEEE Biometrics Council.  She has volunteered for numerous organizations including the FIDO Alliance and Identity Theft Resource Center (ITRC).
\end{IEEEbiography}

\begin{IEEEbiography}[{\includegraphics[width=1in,height=1.25in,clip,keepaspectratio]{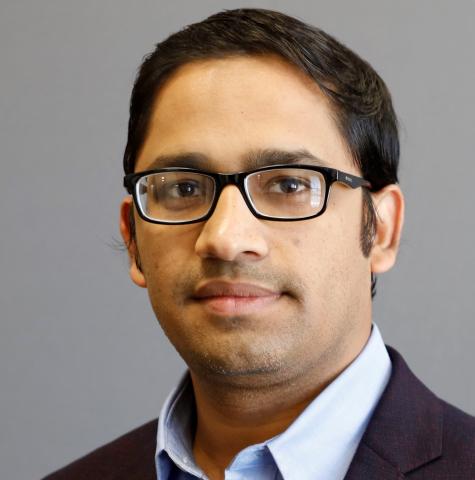}}]{Masudul H Imtiaz}   is currently an assistant professor with the Department of Electrical and Computer Engineering, Clarkson University, Potsdam, NY, USA, and head of the AI Vision, Health, Biometrics, and Applied Computing (AVHBAC) lab. Dr. Imtiaz received bachelor’s and master’s degrees in applied physics, electronics, and communication engineering from the University of Dhaka, Bangladesh, and a Ph.D. degree from the University of Alabama in the summer of 2019. He was a Postdoctoral Fellow with the Department of Electrical and Computer Engineering at the University of Alabama. His research interests include the development of wearable systems, mHealth, deep learning on wearables, biomedical signal processing, and computational intelligence for preventive, diagnostic, and assistive technology. 

\end{IEEEbiography}

\EOD

\end{document}